# $Sr_3Co_2O_6$: a possible frustrated spin-chain material

X. X. Wang,[1,2] J. J. Li,[1,2] Y. G. Shi,[1,3] Y. Tsujimoto,[4] Y. F. Guo,[4] S. B. Zhang,[4] Y. Matsushita,[5] M. Tanaka,[5] Y. Katsuya,[6] K. Kobayashi,[5] K. Yamaura,[1,2,3,*] E. Takayama-Muromachi [2,3,4]

[1] Superconducting Materials Center, National Institute for Materials Science, 1-1 Namiki, Tsukuba, Ibaraki 305-0044, Japan

[2] Department of Chemistry, Graduate School of Science, Hokkaido University, Sapporo, Hokkaido 060-0810, Japan

[3] JST, Transformative Research-Project on Iron Pnictides (TRIP), 1-1 Namiki, Tsukuba, Ibaraki 305-0044, Japan

[4] International Center for Materials Nanoarchitectonics (MANA), National Institute for Materials Science, 1-1 Namiki, Tsukuba, Ibaraki 305-0044, Japan305-0044, Japan

[5] NIMS Beamline Station at SPring-8, National Institute for Materials Science, 1-1-1 Kouto, Sayo-cho, Sayo-gun, Hyogo 679-5148, Japan

[6] SPring-8 Service Co. Ltd., 1-1-1 Kouto, Sayo-cho, Sayo-gun, Hyogo 679-5148, Japan

A dense phase of a post-layered perovskite $Sr_3Co_2O_6$ was created by a high-pressure and high-temperature synthetic route and quenched to ambient conditions. The structure of $Sr_3Co_2O_6$ was determined by synchrotron X-ray powder diffraction. $Sr_3Co_2O_6$ crystallizes in the $K_4CdCl_6$-type structure (an infinite chain-type structure), which is 6% denser than the known, iso-compositional layered perovskite $Sr_3Co_2O_{7-\delta}$ ($\delta \sim 1$). The $K_4CdCl_6$-type $Sr_3Co_2O_6$ shows complex magnetic behavior reflecting the spin-chain nature as well as the analogue oxides $Ca_3Co_2O_6$ and $Ca_3CoRhO_6$, whose microscopic magnetic origin is still highly debated.

Post-perovskite and post-spinel transitions have received increasing attention to better understand the nature of the earth's mantle and in addition, they are expected to yield valuable insights for materials science [1-9]. The post-perovskite and post-spinel materials are usually few % denser than the iso-compositional perovskite and spinel materials, respectively. The denser phases have anisotropic structures, tending either toward two dimensional (post-perovskite) [2,7] or one dimensional (post-spinel) [6,9] in contrast to the perovskite and spinel phases, which are structurally isotropic; hence one might anticipate observing unusual electronic states resulting from the anisotropy that may appear when crossing the transition. For instance, the perovskite oxides $CaRhO_3$ and $CaRuO_3$ undergo a structure transition upon high-pressure heating and, consequently, exhibit significantly different magnetic and electrical properties [1,2,10,11]. Similarly, the spinel oxide $LiMn_2O_4$ undergoes the post-spinel transition via heating and exhibits unique magnetic and electrical behavior [6]. For these reasons, investigations of this class of oxides is of interest both for materials science as well as for the geological science and, in fact, studies aimed at developing materials that exhibit unusual superconductivity via carrier doping of the anisotropic electronic states are ongoing.

Herein we report the successful synthesis of a denser phase of $Sr_3Co_2O_6$ via high-pressure and high-temperature heating. Synchrotron X-ray diffraction experiments confirmed that the trivalent $Co^{3+}$ ($3d^6$) compound crystallizes in the $K_4CdCl_6$-structure type rather than in the Ruddlesden-Popper-type structure (also known as "the layered perovskite-type"), usually observed for $Sr_3Co_2O_{7-\delta}$ ($0 < \delta < \sim1$) [12-14]. The $K_4CdCl_6$-type $Sr_3Co_2O_6$ is structurally anisotropic, consisting of infinite magnetic chains consisting of alternating distorted $CoO_6$ octahedra and $CoO_6$ trigonal prisms that share common faces. These chains are separated from each other by strontium cations [15,16]. The novel $K_4CdCl_6$-type $Sr_3Co_2O_6$ is 6.0% denser than the layered perovskite $Sr_3Co_2O_{7-\delta}$ ($\delta\sim1$), and it is thus warranted to call this novel denser phase a "post-layered perovskite" phase, by comparison with other known post-perovskites [7].

Intriguingly, the post-layered perovskite $Sr_3Co_2O_6$ exhibits possible frustrated magnetism that is accompanied by a magnetization plateau.  The complex magnetic behavior is highly reminiscent of what was observed for the isostructural spin-chain Co oxides $Ca_3CoMO_6$ ($M$ = Co, Rh, Ir), those microscopic magnetic origin is still highly debated [17-29].  Compared with the magnetic properties of $Ca_3Co_2O_6$, the characteristic temperatures (often labeled as $T_{c1}$ and $T_{c2}$) and the magnitude and width of the magnetization plateau are almost unchanged, although the unit-cell volume is expanded 12.4% due to the presence of strontium rather than calcium between the cobalt oxide chains.  This is surprising because magnetic frustration is usually sensitive to arrangement of spins and the result of a delicate balance of multiple magnetic interactions [30].  As far as we know there are very few materials exhibiting robust magnetic frustration over a large volume change.  The novel spin-chain compound $Sr_3Co_2O_6$ thus provides opportunities to shed more light on the microscopic origin of the frustrated magnetism, which is still highly debated, observed for the spin-chain oxides $Ca_3CoMO_6$.  In this paper, we present the synthesis and the frustrated magnetic properties of the $K_4CdCl_6$-type $Sr_3Co_2O_6$.

The polycrystalline post-layered perovskite $Sr_3Co_2O_6$ was prepared by heating a stoichiometric mixture of Co powder (99.5%, Wako) and laboratory made $SrO_2$ [31] at a pressure of 6 GPa and 1400 ºC for 1 hr in a belt-type high-pressure apparatus.  A Pt capsule was used to seal the mixture.  The sample was characterized by synchrotron X-ray powder diffraction using a large Debye-Scherrer camera at the BL15XU beam line of SPring-8 (SXRD, $\lambda$ = 0.65297 Å) [32].  The oxygen content was measured in thermo-gravimetric analysis using hydrogen (5%) and nitrogen mixture gas; 6.08 per the formula unit (f.u.) was obtained, confirming a stoichiometric (or a nearly stoichiometric) oxygen composition.  The temperature dependence of the dc magnetic susceptibility ($\chi$) was measured in a SQUID magnetometer (Quantum Design, MPMS) between 2 K and 400 K in an applied magnetic field of 1 kOe under zero-field cooling (ZFC) and field cooling (FC) conditions.  The isothermal magnetization (*M*) was measured between -70 and 70 kOe and ac magnetic susceptibility ($\chi_{ac}$) was collected using an ac-field amplitude of 5 Oe. For the magnetic measurements, fine powder of $Sr_3Co_2O_6$ was loosely gathered into a holder.

Structural parameters were refined by a Rietveld analysis using the program RIETAN [33]; where the structure of $Ca_3Co_2O_6$ was used as a starting model. The final structure parameters and the refined pattern are shown in Table 1 and Fig. 1, respectively. Since the *R* values are reasonably small and the obtained parameters are comparable with those of $Ca_3Co_2O_6$ [15], we can reasonably conclude that the structure model for $Ca_3Co_2O_6$ is applicable to this compound. The structure of $Sr_3Co_2O_6$ (space group *R*-3*c*) appears one-dimensional, Fig. 1. In the structure, $Co1O_6$ octahedra and $Co2O_6$ trigonal prisms share faces along the c-axis, forming infinite chains that are positioned in a triangular arrangement. To confirm the likely oxidation states of the cobalt atoms in the octahedra and trigonal prisms, we calculated the bond-valence sum of all the metal atoms [34]; 3.15 (Co1), 2.08 (Co2), and 2.12 (Sr). The values are close to those found for cobalt in $Ca_3Co_2O_6$; 3.34 (Co1), 2.22 (Co2), and 2.16 (Ca) [15], underscoring the structurally very similar coordination environments of the cobalt atoms in the two structures. The bond valence sums for Co are expected to be close to 3, where the observed deviation may be due to the trigonal distortion of the chain structure as discussed in Ref. 24. The lattice parameters of the related Co oxides are compared in Table 2.

Fig. 2a shows the $\chi$ vs. *T* and $\chi^{-1}$ vs. *T* (inset) data measured at 1 kOe for the post-layered perovskite $Sr_3Co_2O_6$. It appears that the susceptibility data follows the Curie-Weiss law above approximately 300 K, as shown by the $\chi^{-1}$ vs. *T* data. Using the spin only expression for the magnetic susceptibility, $\chi(T) = N_A\mu^2_{eff}/3\pi(T-\Theta_W)$, where $N_A$ is the Avogadro constant and $\Theta_W$ is the Weiss temperature, the linear portion of the inverse susceptibility data was fit. The effective magnetic moment ($\mu_{eff}$) was calculated to be 6.14(6) $\mu_B$ per f.u., which compares well with the observed moment of 6.08 $\mu_B$ for $Ca_3Co_2O_6$ [26]. To obtain the calculated value of 6.1 $\mu_B$ per f.u. a Landé factor, *g* = 2.5, was used. This suggests that the spin-orbital coupling of Co is very large. In fact, a large orbital moment of 1.7 $\mu_B$ for Co2 in $Ca_3Co_2O_6$ was found by an X-ray magnetic circular dichroism study [24]. The large $g$ value may be responsible for the highly Ising-like features of the Co spins [25]. The Weiss constant, $\Theta_W$ was estimated to be +85(5) K, indicating that the major magnetic interaction is ferromagnetic (FM). Surprisingly, this exactly matches with the value reported for $Ca_3Co_2O_6$ [28].

In the $\chi$ vs. $T$ plot, at least two anomalies are obvious at ~22 K and ~9 K, that, phenomenologically, correlate well with $T_{c1}$ and $T_{c2}$ of $Ca_3Co_2O_6$. In order to investigate those anomalies, the ac susceptibility, $\chi_{ac}$ (= $\chi'$ + $i\chi''$), was measured in the vicinity of $T_{c1}$ and $T_{c2}$. As shown in Figs. 2b and 2c, they can be clearly identified as a magnetically glassy feature, by the frequency dependent peak-shift at $~T_{c2}$. The characteristic constant $K$ (= $\Delta T_f/T_{c2}/\Delta \log f$) estimated from the peak shift is 0.22, which is close to the $K$-value found for $Ca_3Co_2O_6$ [28], and for the super-paramagnetic compound $\alpha$-$(Ho_2O_3)(B_2O_3)$ [30], and is much larger than values (<0.1) of conventional spin-glass metal alloys such as AuFe [30]. This indicates that the glassy feature is not like a conventional spin-glass, but more like a cluster glass under the influence of increasing frustration as discussed for $Ca_3Co_2O_6$ [27].

Fig. 3 shows thermal evolution of the magnetization curves between 2 K and 22 K, showing the appearance of a magnetization plateau at approximately one third of the full magnetization. The evolution of this plateau seems to be coupled to the sequential magnetic transition at $T_{c1}$ and $T_{c2}$. The magnitude of the saturated moment of 4.5 $\mu_B$ per f.u. is comparable with the theoretical value of 5.0 $\mu_B$ per f.u. ($g$ = 2.5). Magnetization kinks at ~0, 10, 22, and 38 kOe (estimated by the peak positions of the differential magnetization curve at 2 K, not shown) are also comparable to those of $Ca_3Co_2O_6$ (~0, 12, 24, 36 kOe) [17]. It should be noted that the present data were obtained from the polycrystalline samples, although the magnetization reaches nearly the expected value. This is surprising because the polycrystalline data of $Ca_3Co_2O_6$ are usually much lower (approximately 70 %) than the single crystalline data because of random orientation of the anisotropic magnetization [26,29,35]. In the present case for $Sr_3Co_2O_6$, the fine powder may be somewhat aligned along the applied magnetic field direction, otherwise degree of the Ising-like magnetic anisotropy becomes somewhat smaller by the unit cell expansion. The spins possibly order without difficulty toward the $H$ direction regardless of the crystal direction. Further study using a single crystalline $Sr_3Co_2O_6$, if it is available, can address the issue.

An analysis of the magnetic data for $Ca_3Co_2O_6$ indicates that the intra-chain magnetic coupling between the neighbor Co2 atoms (often labeled as $J_1$) is ferromagnetic (FM) while the inter-chain magnetic couplings between the neighbor Co2 atoms ($J_2$) and between the next nearest neighbor Co2 atoms ($J_3$) are antiferromagnetic (AF). Absolute values of $J/k_B$ are estimated to be ~25 K, <1 K, <1 K, for $J_1$, $J_2$ and $J_3$, respectively [18]. Recent findings question the Ising-like picture for Co2 spins in $Ca_3Co_2O_6$ [19,20], which is commonly mentioned in early reports [21-29]. The short range intra-chain FM correlations strengthen on cooling and the FM chains exhibit partial long range order and couple antiferromagnetically with each other at $T_{c1}$. The partially disordered AF state is modulated along c-axis [20]. Upon additional cooling, the disordered spins freeze at $T_{c2}$ in a frustrated state [27]. The complex magnetization curve that exhibits the 1/3 magnetization plateau is reproduced to some extent by theoretical models, which take account of the magnetic frustration [21-23,29]. Overall, there are little experimental contradictions among the investigations so-far. Due to the close similarity between the magnetic properties of $Sr_3Co_2O_6$ and $Ca_3Co_2O_6$ (Table 2), it is likely that these two materials share a common magnetic structure, in spite of the 12.4% unit-cell volume difference.

The increase in the distance between the intra-chain neighbor Co2 atoms is 2.2%, and between the inter-chain neighbor Co2 atoms is 0.92%, and between the inter-chain next neighbor Co2 atoms is 4.11%. Thus, the structure accommodates the larger strontium cation by anisotropically expanding the chains, resulting in an impaired balance of FM and AF interactions, including $J_1$ to $J_3$. The present data, however, demonstrate that the frustrated magnetic nature is not altered significantly in spite of the structural expansion. Future quantitative analyses by first principle calculations may help to ultimately decide if indeed the present results agree with our current understanding of the magnetism of $Ca_3Co_2O_6$.

Finally, the post-layered perovskite phase $Sr_3Co_2O_6$ is 6.0% denser than the layered perovskite phase, indicating that the high-pressure heating facilitates the structure transition. The increase in the density is not far from that of the post-perovskite transition (~2%) [1,2,7] and the post-spinel transition (~6-10 %) [6,9]. Crossing the post-layered perovskite transition may lead to varieties of magnetically and electrically active compounds, if it can be extended to other layered perovskite compounds. As far as we know, there are very few studies of phases denser than the layered perovskite [36].

The observed spin-chain magnetism of $Sr_3Co_2O_6$ is reminiscent of what was observed for $Ca_3Co_2O_6$, even quantitatively, with respect to the values of $T_{c1}$, $T_{c2}$, $\mu_{eff}$, $\Theta_W$, $K$, and the width of the 1/3 magnetization plateau, suggesting that the peculiar magnetism in both phases can be characterized by the same magnetic model. Moreover, the present results open a potential route to ways to dope electrical carriers into $A_3Co_2O_6$ ($A$ = Ca and Sr). Since carrier doping into $Sr_3Co_2O_6$ to affect the intriguing quantum magnetic state can potentially result in unexpected correlated properties, we are actively pursuing high pressure experiments to establish the reaction conditions that will result in doping.

We thank Y. Shirako, M. Miyakawa, and H.-C. zur Loye for helpful discussions. This research was supported in part by "WPI Initiative on Materials Nanoarchitectonics" from the Ministry of Education, Culture, Sports, Science and Technology, Japan; "Research Seeds Quest Program," managed by the Japan Science and Technology Agency; and "Grants-in-Aid for Scientific Research (20360012, 21540330, 22246083)" and "Funding Program for World-Leading Innovative R&D on Science and Technology (FIRST Program)," managed by the Japan Society for the Promotion of Science.

Table 1 Structure and isotropic displacement parameters of the post-layered perovskite $Sr_3Co_2O_6$.[a]

| Atom | site | $x$ | $y$ | $z$ | $g$ | $B_{iso}$(Å$^2$) |
|---|---|---|---|---|---|---|
| Sr | 18$e$ | 0.36953(4) | 0 | 0.25 | 1 | 0.58(1) |
| Co1 | 6$b$ | 0 | 0 | 0 | 1 | 0.27(3) |
| Co2 | 6$a$ | 0 | 0 | 0.25 | 1 | 0.67(3) |
| O | 36$f$ | 0.1676(3) | 0.0250(3) | 0.1143(2) | 1 | 0.22(6) |

[a] Note: Space group $R\text{-}3c$ (No. 167); $Z = 6$ (hexagonal setting); $a = 9.61107(4)$ Å, $c = 10.70110(5)$ Å, $V = 856.056(7)$ Å$^3$; $R_{wp} = 1.481$ %, $R_p = 1.031$ %. $d$(Co1-O) = 1.9389 Å (×6), $d$(Co1-Co2) = 2.6753 Å (×2), $d$(Co2-O) = 2.0917 Å (×6).

Table 2 Comparison of the structure and the magnetic parameters of the post-layered perovskite $Sr_3Co_2O_6$ and isostructural compounds.

| Compounds | Lattice parameters | | | Magnetic properties | | | | | | Ref. |
|---|---|---|---|---|---|---|---|---|---|---|
| | $a$/Å | $c$/Å | $V/V_{Ca}$ | $T_{c1}$ (K) | $T_{c2}$ (K) | $\Theta_W$ (K) | $\mu_{eff}$ ($\mu_B$/f.u.) | $M_{sat}$ ($\mu_B$/f.u.) | $K$ | |
| $Sr_3Co_2O_6$ ($R\text{-}3c$) | 9.611 | 10.701 | 1.12 | 22 | 9 | 85 | 6.14 | 4.5 | 0.22 | This work |
| $Ca_3Co_2O_6$ ($R\text{-}3c$) | 9.079 | 10.381 | 1 | 24 | 12 | 80 | 6.08 | 4.8 | 0.17 | 17,28 |
| $Ca_3CoRhO_6$ ($R\text{-}3c$) | 9.202 | 10.730 | 1.06 | 90 | 25 | 150 | 4.44 | 4 | 0.10 | 29 |

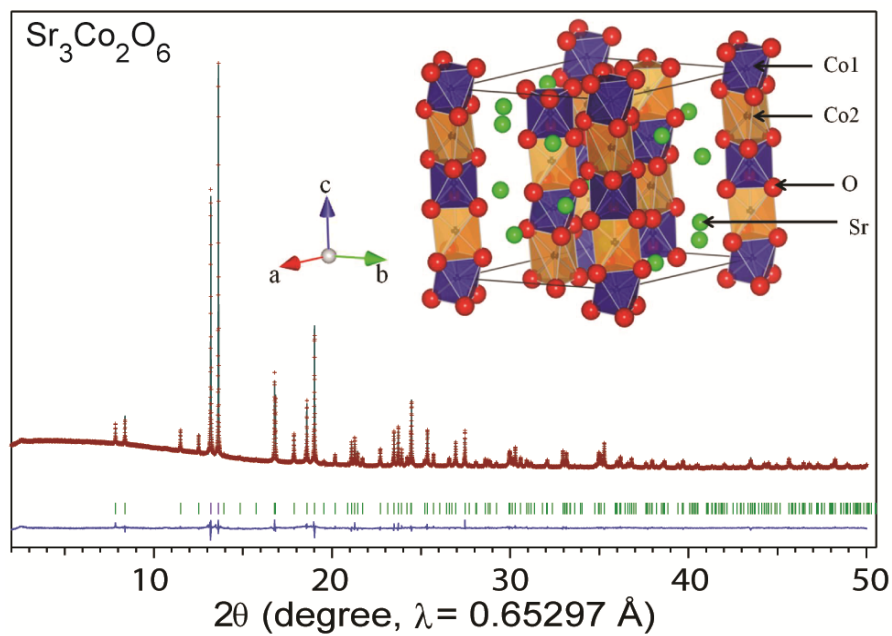

Fig. 1 The Rietveld analysis of the SXRD profile for the post-layered perovskite $Sr_3Co_2O_6$. Cross markers and solid lines show the observed and calculated profiles, respectively, and the difference is shown at the bottom. The positions of Bragg reflections are marked by ticks. (Inset) Sketch of the crystal structure of the post-layered perovskite $Sr_3Co_2O_6$.

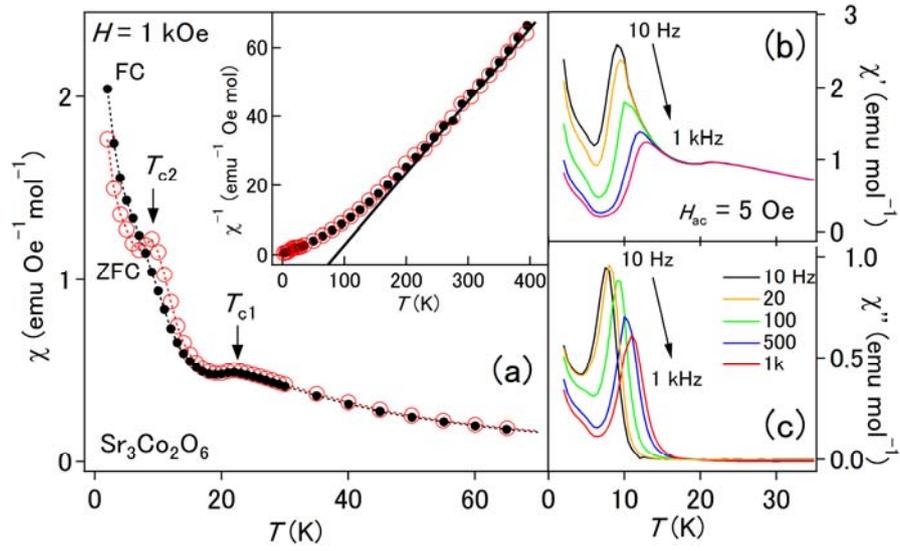

Fig. 2 (a) $T$ dependence of $\chi$ of the polycrystalline post-layered perovskite $Sr_3Co_2O_6$. Inset shows an inverse plot of the $\chi$ data. Solid line corresponds to the fit. (b,c) $T$ and frequency dependence of ac-$\chi$ (= $\chi'$ + $i\chi''$) of $Sr_3Co_2O_6$.

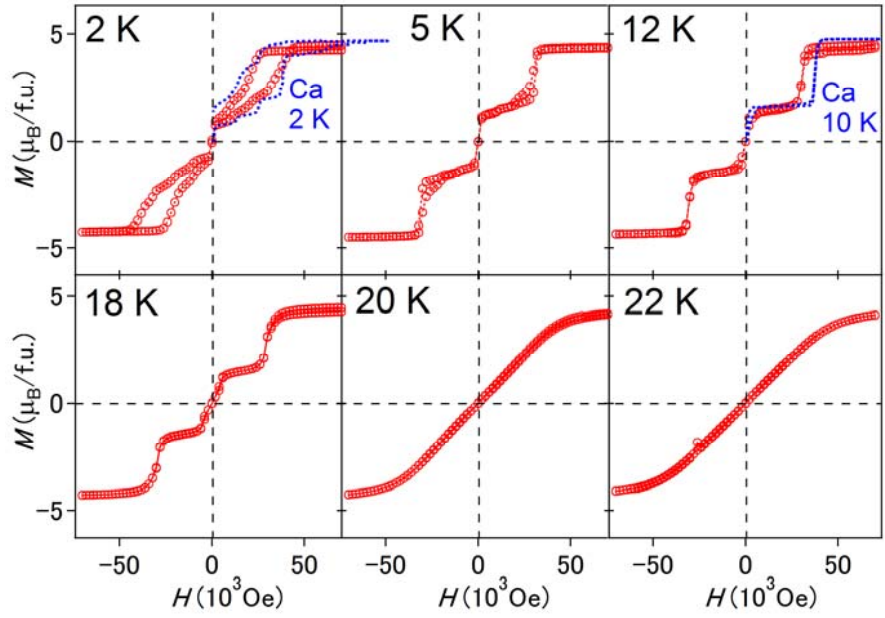

Fig. 3  Isothermal magnetization of the polycrystalline post-layered perovskite $Sr_3Co_2O_6$. The dotted curves are for the single crystalline $Ca_3Co_2O_6$ ($H$ //c) taken from Ref. 26.